\documentclass[sigconf]{acmart}

\AtBeginDocument{%
  }

\setcopyright{acmlicensed}
\copyrightyear{2025}
\acmYear{2025}
\acmDOI{XXXXXXX.XXXXXXX}

\acmISBN{978-1-4503-XXXX-X/18/06}

\usepackage{multirow,tabularx}
\usepackage{amsmath} 
\usepackage{tcolorbox}
\usepackage{enumitem}
\usepackage{algorithm}  
\usepackage{algorithmicx}
\usepackage{algpseudocode} 
\usepackage{subcaption}
\usepackage{amsfonts} 
\usepackage{bm}

\newcommand{\ie}{\emph{i.e., }}
\newcommand{\eg}{\emph{e.g., }}

\newcommand{\aka}{\emph{aka. }}

\begin{document}

\title{Generate and Instantiate What You Prefer: Text-Guided Diffusion for Sequential Recommendation}


\author{Guoqing Hu}
\authornote{Both authors contributed equally to this work.}
\affiliation{%
  \institution{University of Science and Technology of China}
  \country{}}
\email{HugoChinn@mail.ustc.edu.cn}

\author{Zhengyi Yang}
\authornotemark[1]
\affiliation{%
  \institution{University of Science and Technology of China}
  \country{}}
\email{yangzhy@mail.ustc.edu.cn}
\author{Zhibo Cai}
\affiliation{%
  \institution{Renmin University of China}
  \country{}}
\email{caizhibo@ruc.edu.cn}

\author{An Zhang}
\affiliation{%
  \institution{National University of Singapore}
  \country{}}
\email{an_zhang@nus.edu.sg}

\author{Xiang Wang}
\affiliation{%
  \institution{University of Science and Technology of China}
  \country{}}
\email{xiangwang1223@gmail.com}




\begin{abstract}
  Recent advancements in generative recommendation systems, particularly in the realm of sequential recommendation tasks, have shown promise in enhancing generalization to new items. Among these approaches, diffusion-based generative recommendation has emerged as an effective tool, leveraging its ability to capture data distributions and generate high-quality samples. 
Despite effectiveness, two primary challenges have been identified: 1) the lack of consistent modeling of data distribution for oracle items; and 2) the difficulty in scaling to more informative control signals beyond historical interactions. 
These issues stem from the uninformative nature of ID embeddings, which necessitate random initialization and limit the incorporation of additional control signals. 
To address these limitations, we propose \textbf{iDreamRec} to involve more concrete prior knowledge to establish item embeddings, particularly through detailed item text descriptions and advanced Text Embedding Models (TEM). More importantly, by converting item descriptions into embeddings aligned with TEM, we enable the integration of intention instructions as control signals to guide the generation of oracle items.
Experimental results on four datasets demonstrate that iDreamRec not only outperforms existing diffusion-based generative recommenders but also facilitates the incorporation of intention instructions for more precise and effective recommendation generation. 
\end{abstract}

 


\maketitle

\section{Introduction}\label{sec:intro}

\begin{figure}[t]
    \centering
    \begin{subfigure}{0.49\linewidth}
        \centering
        \includegraphics[width=0.9\linewidth]{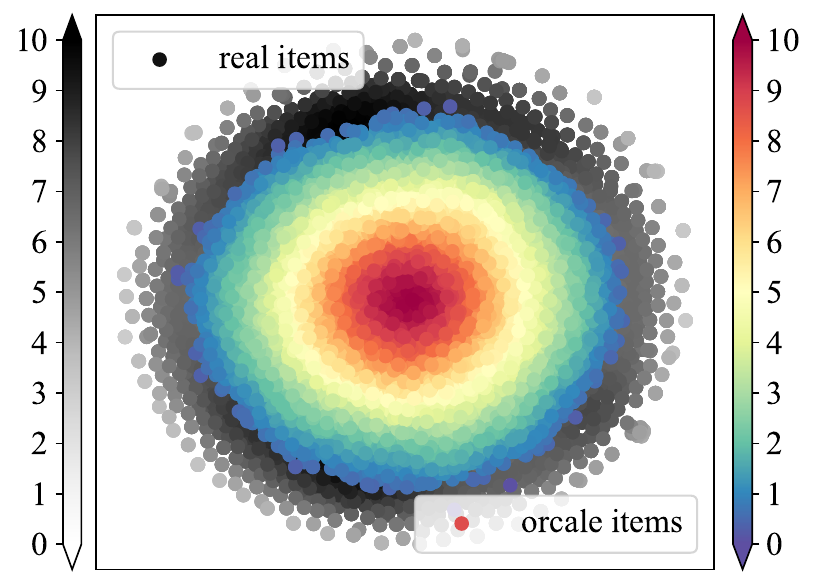}
        \caption{no guidance}
        \label{uncond}
    \end{subfigure}
    \begin{subfigure}{0.49\linewidth}
        \centering
        \includegraphics[width=0.9\linewidth]{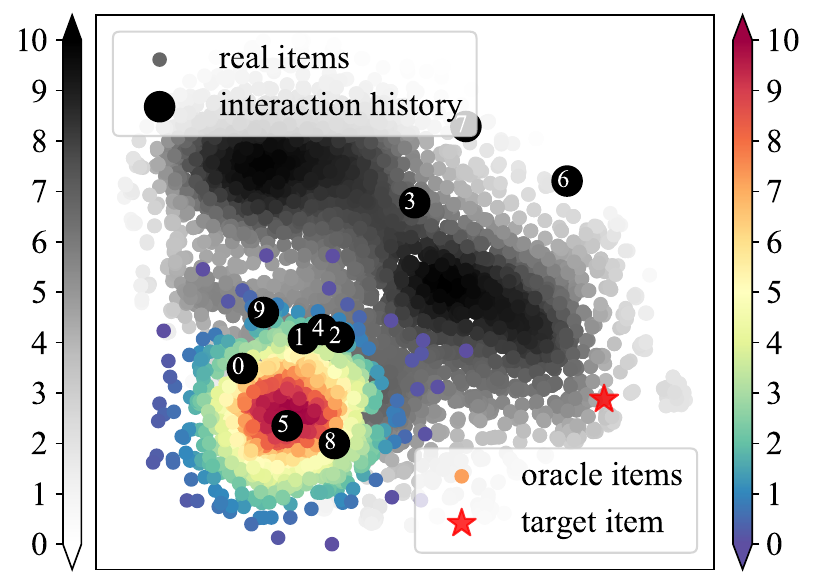}
        \caption{interaction history}
        \label{history}
    \end{subfigure}
    \qquad
	
    \begin{subfigure}{0.49\linewidth}
		\centering
		\includegraphics[width=0.9\linewidth]{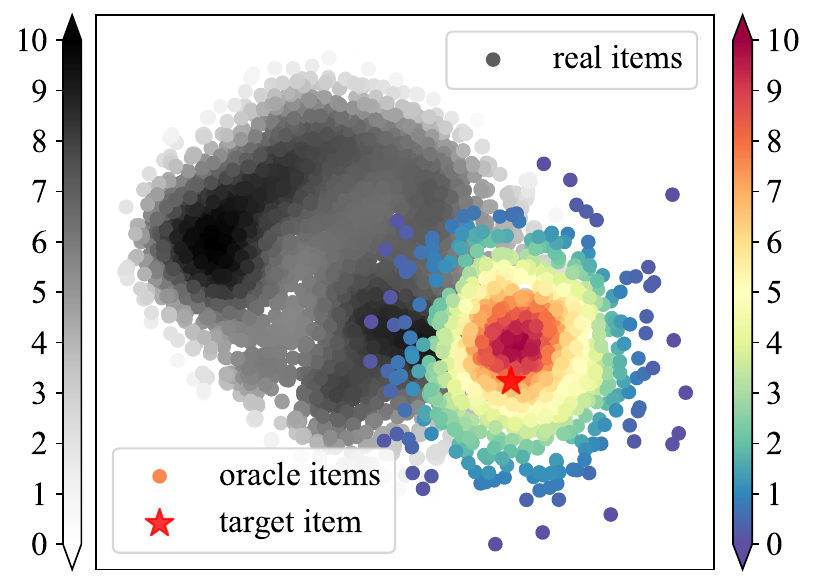}
		\caption{intention}
		\label{intention}
    \end{subfigure}
    \begin{subfigure}{0.49\linewidth}
		\centering
		\includegraphics[width=0.9\linewidth]{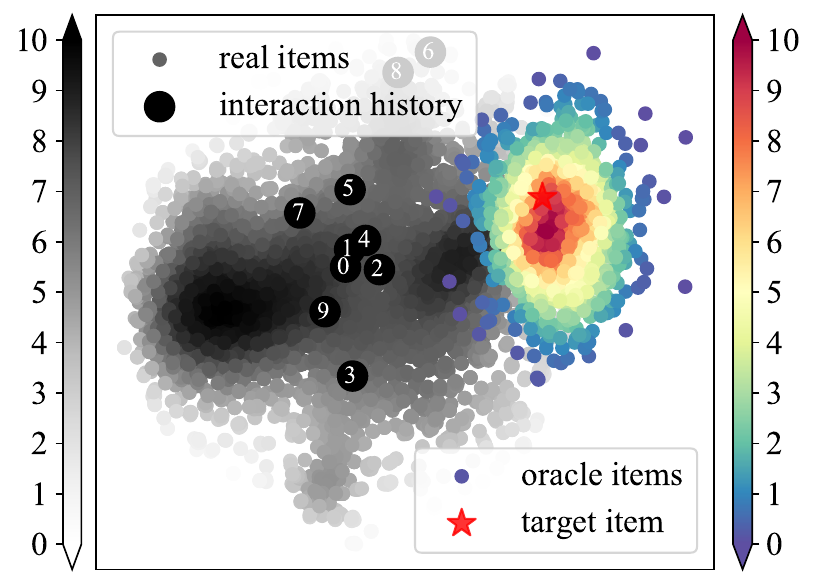}
		\caption{both}
		\label{both}
    \end{subfigure}
    \caption{The t-SNE visualization of item embeddings under different guidance signals. The color bar scale represents the relative density. When the oracle items generated based on the user's interaction history do not align with the user's interests, introducing user intention control signals can improve the recommendation effectiveness.}
    \label{both}
\end{figure}


Recent advances in generative recommendation, especially in sequential recommendation scenarios \cite{SASRec, Caser, GRURec}, have shown promising results due to the ability to generalize better to new items \cite{TIGER, DreamRec, GenRec-Survey, DRM, LD4MRec}.
Among various generative frameworks, diffusion models \cite{DDPM, iDDPM, DDIM, SBM, iSBM, SDE, Consistency-model} have emerged as highly effective,  largely due to their theoretical strengths in modeling complex data distributions and generating high-quality samples.
For example, DreamRec \cite{DreamRec} employs a diffusion model, conditioned on the interaction histories of users, to predict the distribution of subsequent items (\ie oracle items) and ground them to the real items users are likely to interact with.
The learning objective in this framework is to generate oracle item embeddings through conditional diffusion processes, where the embeddings are simultaneously optimized alongside the model parameters. 

Despite the effectiveness of DreamRec, we identify two inherent limitations that constrain its potential to fully exploit diffusion models in generative recommendation:
\begin{itemize}[leftmargin=*]
    \item \textbf{Inconsistent Data Distribution of Oracle Items.}
    While diffusion models, in general, are excel in capturing consistent data distributions according to their theoretical underpinnings \cite{DDPM, SBM, SDE}, DreamRec diverges from this principle.
    Instead of relying on pre-existing fixed item representations, DreamRec randomly initializes item embeddings which are then jointly optimized with model parameters during training.
    This process lacks a stable anchor in the latent space, causing item embeddings to fluctuate and evolve over time.
    Consequently, this instability leads to inconsistent data distribution for oracle items, ultimately degrading the performance of generative recommendation.
    
    \item \textbf{Limited Exploration to More Informative Controls.}
     Although DreamRec effectively uses interaction histories as control signals to generate oracle items, it struggles to incorporate richer control signals, such as user intentions.
     For example, a user might express an intent like ``I want to read a book about a young boy who discovers his magical heritage, attends a hidden school of wizardry, and confronts a dark evil'' (the example of intention in Figure \ref{intention}), which could lead to a recommendation like Harry Potter.
     However, as illustrated in Figure \ref{history}, the reliance on interaction histories limits the model’s ability to generate more nuanced and relevant recommendations, thereby constraining the full generative potential of diffusion models.
\end{itemize}

We attribute these limitations to a core issue: the reliance on learning item embeddings based solely on IDs.
This not only requires random initialization, but also makes it challenging to incorporate additional control signals.
A promising solution is to enrich item embeddings with more substantial prior knowledge beyond mere ID information.
Furthermore, if the prior knowledge is interpretable by humans, it could enable the integration of intention instructions alongside interaction histories to better guide the generation process.
Inspired by the capabilities of Large Language Models (LLMs) \cite{GPT3,LLAMA},  leveraging the extensive world knowledge embedded within LLMs presents a valuable opportunity.
However, how to effectively apply LLMs to create informative prior knowledge for item embeddings remains largely unexplored in diffusion-based generative recommendation.
In this paper, we propose \textbf{iDreamRec} (\textbf{i}ntention-guided \textbf{DreamRec}), incorporating text embeddings from LLMs to accurately model item distributions and generate oracle items.
Specifically, given the metadata about items (\eg titles), we first prompt GPT \cite{GPT3} to generate detailed textual descriptions, offering richer content than simple item IDs.
To transform such descriptions into item embeddings, we employ the advanced Text Embedding Models (TEM) \cite{GTE, RoBERTa,  SimCSE, SGPT} provided by OpenAI \footnote{\url{https://platform.openai.com/docs/guides/embeddings}}.
This approach allows for converting variable-length descriptions into consistent, context-aware item embeddings.
On this consistent embedding space, we train a diffusion model conditioned on interaction histories using a classifier-free guidance strategy \cite{CFG} as Algorithm \ref{alg:train}.
Crucially, as item embeddings align with TEM, intention instructions can also be translated into embeddings within the same space, providing refined control signals for generating oracle items (as shown in Figure \ref{intention}).
Intent instructions can also be integrated with interaction histories to generate more nuanced and relevant oracle items (as illustrated in Figure \ref{both}). 
For concrete examples, please refer to Figure \ref{figCase}.

Evaluations on four benchmark datasets suggest that {iDreamRec} not only outperforms current diffusion-based recommenders (\eg DreamRec \cite{DreamRec}, DiffRec \cite{DRM}), but also enables the integration of intention instructions to guide the generation of oracle items.

\section{Related Work}


\textbf{Diffusion models} have shown extraordinary performance in various generative tasks, including image synthesis \cite{CG, DiT}, text generation \cite{DiffText1,DiffText2}, 3D molecular modeling \cite{DiffMolecule}, biological sequence design \cite{DirichletDiff4BioSeq} and anomaly detection \cite{DiffAnomaly}. 
The effectiveness of diffusion models stems from their training stability and high sampling quality in capturing the data distribution given observed samples compared to GANs \cite{GAN1, GAN2} and VAEs \cite{VAE1, VAE2}. 
In general, diffusion models can be derived through the following three approaches: denoising diffusion probabilistic models (DDPM) \cite{DDPM,DDIM}, score-based generative models (SGM) \cite{SBM, iSBM}, and stochastic differential equations (SDE) \cite{SDE}, which has theoretically shown to be equivalent and achieve similar performance in practice \cite{SDE, 3DiffEqual}. 
Specifically, SGM is a discrete form of variance-exploding SDE, while DDPM is a discrete form of variance-preserving SDE. 
In terms of application, works tend to adopt the DDPM framework (\eg StableDiffusion \cite{LDM}, DiT \cite{DiT}) due to its flexibility in adapting to different scenarios.


\textbf{Diffusion-based recommenders} can be roughly classified into generative and discriminative ones according to their task formulation.
Specifically, diffusion-based discriminative recommenders either predict interaction probability vectors for all items per user \cite{DRM, BinomialDiff4Rec} 
or treat the diffusion model as a powerful enhancement for traditional discriminative recommenders, which is jointly trained under classification or ranking losses such as \textit{cross-entropy} of \textit{Bayesian Personalized Ranking} \cite{Diff4POIRec, DiffuRec, CDD4Rec, SeqDiff}.
To fully leverage the powerful generative capability of the diffusion model, the diffusion-based generative recommender (\ie DreamRec) reshapes sequential recommendation as generation of the subsequent item embedding (\aka oracle item embedding) under the condition of the interaction history \cite{DreamRec}.
However, DreamRec adopts randomly initialized and trainable ID embeddings similar to traditional sequential recommenders \cite{SASRec, Caser, GRURec}, which might bring practical challenges (as discussed in Section \ref{sec:intro}).


\textbf{Text embedding model} (TEM) aims to encode massive text into a unified semantic space with context awareness.
Most TEMs are fine-tuned on large language models (LLM, \eg BERT \cite{BERT}, T5 \cite{T5}, GPT \cite{GPT3}) with the contrastive learning objective. 
In the previous, BERT-based TEMs are widely used to acquire text embeddings for downstream tasks \cite{GTE, RoBERTa, Longformer, SimCSE}.
Recently, as GPT \cite{GPT3} has shown prevailing capability in natural language tasks, it becomes more promising to employ GPT-based TEMs, which are available via OpenAI Embedding API \cite{OpenAI_ada}.

\textbf{Large language models for recommendation} (LLM4Recs) have demonstrated remarkable capabilities. In general, LLM4Recs can be divided into LLM-based recommenders and LLM-enhanced recommenders. LLM-based recommenders consider LLM as the backbone of recommenders \cite{TallRec, BigRec, E4SRec, LLaRA, HSTU}, while LLM-enhanced recommenders view LLM as a powerful feature extractor \cite{RLMRec, LLMRec, AlphaRec}. Despite their powerful performance, LLM-based recommenders still face challenges such as high fine-tuning costs and relatively low recommendation efficiency. In practice, LLM-enhanced recommenders are more lightweight and feasible.
\section{Preliminary}
In this section, we introduce the diffusion-based generative recommendation, basically following DreamRec \cite{DreamRec}.
Here, denoising diffusion implicit models (DDIM) \cite{DDIM} form the core of the diffusion framework. In contrast to the denoising diffusion probabilistic models (DDPM) \cite{DDPM} originally employed in DreamRec, DDIM provides faster sampling and enhances sampling quality.
See Appendix \ref{DDIM} for more details on DDIM and Table \ref{tab:symbols} for the symbol definitions. 

\vspace{5pt}
\noindent\textbf{Sequential recommendation formulation:}
Let $\mathcal{I}$ be the set of all items. 
A interaction history is represented as $\mathbf{v}_{<L} = \left[{v}_1, {v}_2, \ldots, {v}_{L-1}\right]$, and ${v}_{L}$ is the target item subsequent to the sequence. 
Let $\mathcal{D} = \{ \left(\mathbf{v}_{<L}, {v}_L\right)_k \}_{k=1}^{|\mathcal{D}|}$ be the training data collecting the pairs of historical sequences and items of interest, and $\mathcal{D}_t = \{ \left(\mathbf{v}_{<L}, {v}_L\right)_k \}_{k=1}^{|\mathcal{D}_t|}$ be the test dataset.
Typically, each item ${v} \in \mathcal{I}$ is initially translated into its corresponding embedding vector $\mathbf{e}\in\mathbb{R}^{d}$.
Then, the interaction histories can be represented as $\mathbf{e}_{<L} = [\mathbf{e}_1, \mathbf{e}_2, \ldots, \mathbf{e}_{L-1}]$. 
Discriminative sequential recommendation \cite{SASRec, GRURec, Caser} aims to classify the item of interest $\mathbf{e}_L$ from the others in $\mathcal{I}$, conditioned on the interaction histories $\mathbf{e}_{<L}$.
In contrast, generative sequential recommendation, as exemplified by DreamRec, aims to approximate the conditional distribution $q(\mathbf{e}_L|\mathbf{e}_{<L})$ with $p_{\theta}(\mathbf{e}_L^0|\mathbf{e}_{<L})$ via guided diffusion and utilizes it to generate oracle items $\mathbf{e}_L^0$.
Here we elaborate on the forward and inference processes of DreamRec.

\vspace{5pt}
\noindent\textbf{Forward process:}
To approximate $q(\mathbf{e}_L|\mathbf{e}_{<L})$ in DDIM, we firstly perturb the subsequent interacted item embedding at varying noise levels with ascending schedule $\left[\alpha_1,\ldots,\alpha_T\right]$:
\begin{equation}
    \label{eq:perturbe}
    q(\mathbf{e}_L^t|\mathbf{e}_{1:L}) := \mathcal{N}(\mathbf{e}_L^t;\sqrt{\alpha_t}\mathbf{e}_L^0,(1-\alpha_t)\mathbf{I}_d),\quad t\in\left[1,\ldots,T\right],
\end{equation}
where $\mathbf{e}_L^0 = \mathbf{e}_L$ and $\mathbf{e}_L^t$ is the noisy embedding at noise level $\alpha_t\in[0,1]$.
DDIM thus operates as a latent variable model of the form:
\begin{equation}
    p_{\theta}(\mathbf{e}_L^0|\mathbf{e}_{<L}) = \int p_{\theta}(\mathbf{e}_L^{0:T}|\mathbf{e}_{<L}) \mathrm{d}\mathbf{e}_L^{1:T},
\end{equation}
\text{where} $p_{\theta}(\mathbf{e}_L^{0:T}|\mathbf{e}_{<L})=p_{\theta}(\mathbf{e}_L^{T}|\mathbf{e}_{<L})\sum\limits_{t=1}^{T}p_{\theta}(\mathbf{e}_L^{t-1}|\mathbf{e}_L^{t},\mathbf{e}_{<L}).$

The parameters $\theta$ are learned to fit the data distribution $q(\mathbf{e}_L^0|\mathbf{e}_{<L})$ by maximizing the evidence lower bound of the log-likelihood $\log p_{\theta}(\mathbf{e}_L^0|\mathbf{e}_{<L})$. This can be simplified as the following loss function (please refer to Appendix \ref{A.1} for details):
\begin{equation}
\label{eg:loss}
    \mathcal{L}=
    \mathbb{E}_{t\sim \mathcal{U}\left[1,T\right]}
    \mathbb{E}_{(\mathbf{e}_{<L},\mathbf{e}_L^0)\sim \mathcal{D}}
    \left[\Vert \hat{\bm{e}}_{\theta}(\mathbf{e}_{L}^t,t,\mathbf{e}_{<L})-\mathbf{e}_{L}^0 \Vert_2^2\right],
\end{equation}
where $\hat{\bm{e}}_{\theta}(\mathbf{e}_{L}^t,t,\mathbf{e}_{<L})$ is the model's output for $\mathbf{e}_{L}^0$ at time step $t$.

To model the unconditional distribution $q(\mathbf{e}_{L}^0)$, we utilize classifier-free guidance strategy \cite{CFG} which models $p_{\theta}(\mathbf{e}_{L}^0|\Phi)$ to fit $q(\mathbf{e}_{L}^0)$. 
$\Phi$ is a trainable vector used as the substitute for conditional guidance during unconditional inference.
The conditional strength can be adjusted by tuning the guidance parameter $w$, leading to the following update (please refer to Appendix \ref{A.3} for details):
\begin{equation}
    \label{eq:cfg}
    \tilde{\bm{e}}_{\theta}(\mathbf{e}_{L}^t,t,\mathbf{e}_{<L}) = (1+w)\hat{\bm{e}}_{\theta}(\mathbf{e}_{L}^t,t,\mathbf{e}_{<L}) - w\hat{\bm{e}}_{\theta}(\mathbf{e}_{L}^t,t,\Phi).
\end{equation}

\vspace{5pt}
\noindent\textbf{Inference process:}
After training phase, we have an approximate one-step denoiser $\tilde{\bm{e}}_{\theta}(\mathbf{e}_{L}^t,t,\mathbf{e}_{<L})$ towards $\mathbf{e}_L^0$, which is not sufficiently accurate for predicting. 
Thus, the multi-step inference phase is needed for more accurate results through gradually denoising (please refer to Appendix \ref{A.2} for details):
\begin{equation}
    \label{eq:DDIM_inverse}
    p(\mathbf{e}_L^s|\mathbf{e}_L^t,\mathbf{e}_{1:L}) := \mathcal{N}(\mathbf{e}_L^s; \sqrt{\alpha_s} \mathbf{e}_L^0 + \sqrt{1-\alpha_s-\sigma_s^2}\frac{\mathbf{e}_L^t-\sqrt{\alpha_t}\mathbf{e}_L^0}{\sqrt{1-\alpha_t}},\sigma_s^2\mathbf{I}).
\end{equation}
We could set $\sigma_s=0$ when the Equation \eqref{eq:DDIM_inverse} equals a deterministic transformation. Note that Equation \eqref{eq:DDIM_inverse} reduces to the original DDPM sampling if we set $\sigma_s^2=\frac{1-\alpha_s}{1-\alpha_t}\left(1-\frac{\alpha_t}{\alpha_s}\right)$ and $s=t-1$.

In DDIM, multi-step denoising can be performed by choosing arbitrary $s < t$ rather than strictly $s=t-1$. 
The denoising process remains stable as the formula is deterministic:
\begin{equation}
    \label{eq:DDIM_sample}
     \hat{\mathbf{e}}_L^s = \sqrt{\alpha_s} \tilde{\bm{e}}_{\theta}(\hat{\mathbf{e}}_L^t,t,\mathbf{e}_{<L}) + \sqrt{1-\alpha_s}\frac{\hat{\mathbf{e}}_L^t-\sqrt{\alpha_t}\tilde{\bm{e}}_{\theta}(\hat{\mathbf{e}}_L^t,t,\mathbf{e}_{<L})}{\sqrt{1-\alpha_t}},
\end{equation}
which begins with $\hat{\mathbf{e}}_{L}^T \sim \mathcal{N}(\mathbf{0}, \mathbf{I}_d)$. 

\vspace{5pt}
\noindent\textbf{Recommend process:} 
After generating the oracle item embedding $\hat{\mathbf{e}}_L^0$ from the distribution $p_{\theta}(\mathbf{e}_L^0|\mathbf{e}_{<L})$ as Equation \eqref{eq:DDIM_sample}, it is unlikely that $\hat{\mathbf{e}}_L^0$ will exactly match any item in the candidate set $\mathcal{I}$.
Thereby, we could ground the oracle items to the real items by simply computing the dot product of $\hat{\mathbf{e}}_L^0$ with all item embeddings in $\mathcal{I}$.
Then, we could recommend the top-$k$ items with the highest dot products.
 \begin{figure*}[t!] 
    \centering 
    \includegraphics[width=\textwidth]{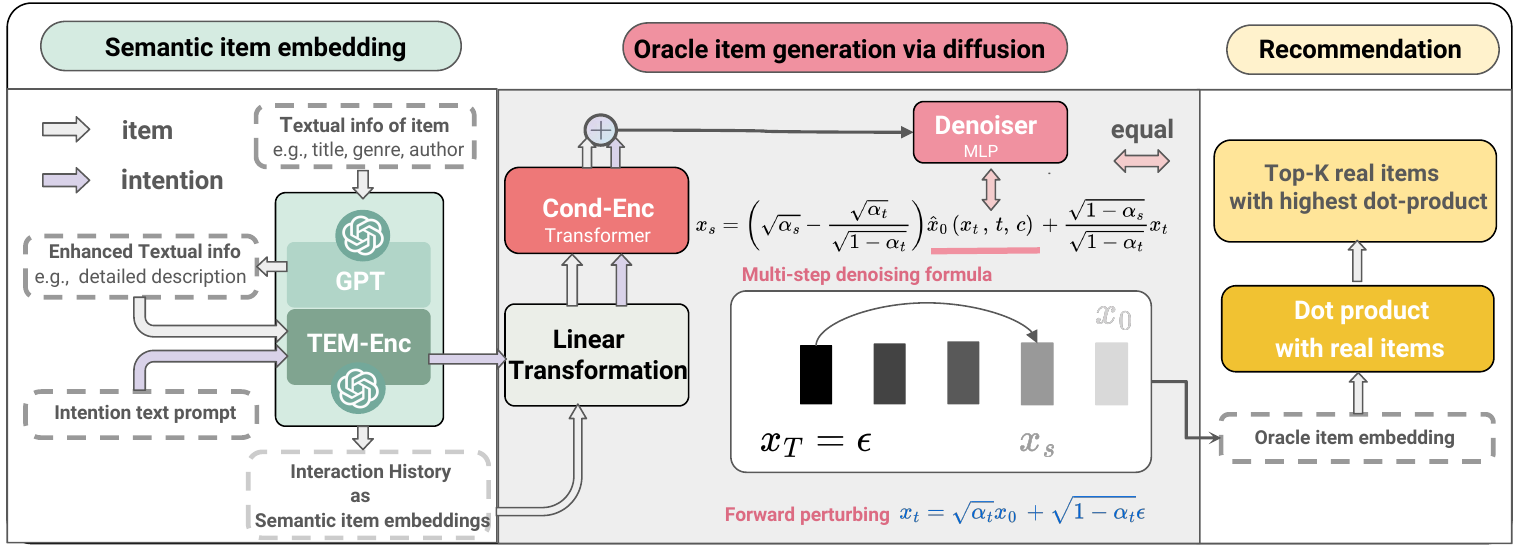}
     \caption{Overall framework of iDreamRec. Firstly, iDreamRec employs Text Embedding Models (TEM) to encode the item description acquired from ChatGPT into unified text embeddings, with variance-preserving transformation to make text embeddings more suitable for diffusion models. 
     Moreover, the generation of iDreamRec can be guided by intention instruction embeddings translated from intention instructions by TEM, allowing to involve more informative control signals. For detailed algorithms, please refer to Appendix \ref{alg}.}
     
     \label{figframework}
 \end{figure*}

\section{Method}


In this section, we introduce our proposed iDreamRec, aiming to build a more consistent item distribution for the diffusion-based generative recommendation, as well as involving intention instructions into the conditional guidance to generate more targeted and relevant oracle items. 

\subsection{Consistent Item Embedding Modeling}

We identify that the core limitation of previous diffusion-based generative recommenders stems from uninformative ID embeddings, which might impair the generalization ability of diffusion models due to inconsistency during training \cite{CDM, CDMT}.
To resolve this issue, we propose incorporating prior knowledge encoded in LLMs to create a more consistent embedding distribution of items for diffusion-based generative recommendation.

\subsubsection{\textbf{Prompting ChatGPT to Enrich Item Description}}

To create a more consistent embedding distribution of items than learnable ID embeddings, it is crucial to include richer prior knowledge about items.
However, in real-world scenarios, the available information about items is often limited.
To overcome this, we propose using ChatGPT \cite{GPT3} to acquire more detailed and coherent descriptions of items, leveraging its extensive world knowledge.
Specifically, given basic information of items $\mathbf{v}$ (\eg titles), we design prompts to extract more comprehensive descriptions from ChatGPT.
Here is an example in the movie recommendation scenario:
\begin{tcolorbox}[boxrule=0.5pt, left=0pt, right=0pt, top=2.5pt, bottom=2.5pt]
        \textit{You are a movie synopsis assistant. For each received movie information in the format of name::genre, You will automatically correct the format of the movie name and output the 
movie synopsis with a similar number of tokens in the format of name::genre::description.}
\end{tcolorbox}
By prompting ChatGPT, we can generate detailed descriptions for item $v$, denoted as $\mathcal{T}_v$:
\begin{tcolorbox}[boxrule=0.5pt, left=0pt, right=0pt, top=2.5pt, bottom=2.5pt]
        \textit{Not Love, Just Frenzy (Más que amor, frenesí) (1996)::Comedy|Drama|Thriller::In the vibrant city of Madrid, a group of friends navigate through a whirlwind of love, friendship, and passion, leading to unexpected consequences.
}
\end{tcolorbox}

\subsubsection{\textbf{Embedding item description via Text Embedding Model}}
 
After enriching the item description (\ie $\mathcal{T}_v$) by prompting GPT, the next step is to construct a consistent item embedding distribution that effectively captures the information within $\mathcal{T}_v$.
A straightforward approach is to leverage open-sourced Language Models (\eg BERT \cite{BERT}, LLaMA \cite{LLAMA}) to generate token-level embeddings of $\mathcal{T}_v$.
However, this presents practical challenges, as the length of descriptions varies significantly across items, which conflicts with the diffusion models' requirement for embeddings to have identical dimensions.
While simple aggregation or pooling methods can generate embeddings with consistent dimensions, they may fail to adequately generalize or retain the necessary information.

Inspired by recent advances in Text Embedding Models (TEMs) \cite{GTE, RoBERTa, Longformer, SimCSE, Sentence-T5, GTR, OpenAI_ada}, which are pre-trained to convert sentences into fixed-dimension embeddings while preserving semantics, we propose leveraging TEMs to generate embeddings for item descriptions.
Specifically, for an item $v$ with a detailed description $\mathcal{T}_v$, we employ a TEM to obtain its corresponding embedding:
\begin{equation}
    \mathbf{e}_{v} = \textbf{TEM-Enc}(\mathcal{T}_v), 
\end{equation}
where $\textbf{TEM-Enc}(\cdot)$ denotes the encoder of TEM, and $\mathbf{e}_{v}$ is the resulting text embedding.
Since $\mathbf{e}_{v}$ encapsulates rich, relevant information from GPT and TEM, it remains frozen throughout training.
Then, we obtain the text embedding matrix $\mathbf{E}\in\mathbb{R}^{|\mathcal{I}| \times d}$ of all items in $\mathcal{I}$, where $d$ is the embedding dimension.
Consequently, diffusion models can focus on capturing a more consistent distribution of items based on GPT-enhanced text embeddings from TEM, rather than relying on uninformative, trainable ID embeddings.

\subsubsection{\textbf{Linear Transformation for Text Embeddings}} \label{Sec:LT}

Inspired by the whitening transformation for text semantic similarity \cite{WhitenBert} (\ie an orthogonal transformation with scaling for text embeddings), we propose a linear transformation $\mathcal{A}(\cdot):\mathcal{A}(\mathbf{e})=(\mathbf{e}-\mathbf{a})\mathbf{A}$ for text embedding matrix $\mathbf{E}$ where $\mathbf{a}\in\mathbb{R}^d$ and $A\in\mathbb{R}^{d\times d}$ both are heuristically defined.\
For example, this transformation can either be a simple scaling constant $a\in \mathbb{R}$ or an orthogonal transformation with scaling, known as a normal transformation. Specifically, $\mathcal{A}(\cdot):\mathcal{A}(\mathbf{e})=a\mathbf{e}$ or $\mathcal{A}(\cdot):\mathcal{A}(\mathbf{e})=(\mathbf{e}-\mathbf{a})\mathbf{O}\mathbf{D}$ where $\mathbf{D}$ is a diagonal matrix and $\mathbf{O}$ is an orthogonal matrix.

To pre-define a normal transformation for text embedding matrix $\mathbf{E}$, we can first calculate its variance matrix $\text{Var}(\mathbf{E})$ and then conduct orthogonal matrix decomposition \cite{LinearA1, LinearA2, PCA}:
\begin{equation}
    \label{eq:Osim}
    \text{Var}(\mathbf{E}) =\mathbf{O} \Lambda \mathbf{O}^T,
\end{equation}
where matrix $\mathbf{O}$ is orthogonal and $\Lambda$ is diagonal. Therefore, various normal transformations can be constructed, such as $\mathcal{A}(\mathbf{e}) = \mathbf{e}\mathbf{O}\Lambda^{-\frac{1}{2}}$, $\mathcal{A}(\mathbf{e}) = \mathbf{e}\mathbf{O}\Lambda^{-\frac{1}{2}}\mathbf{O}$, and $\mathcal{A}(\mathbf{e}) = \mathbf{e}\mathbf{O}\Lambda^{-\frac{1}{2}}\mathbf{O}^T$. Here, $\Lambda^{-\frac{1}{2}}$ denotes the reciprocal square root of the diagonal elements of $\Lambda$.

Inheriting from whitening transformation, our linear transformation $\mathcal{A}(\cdot)$ can enhance the effectiveness of semantic matching for text embeddings \cite{WhitenBert}. 
Beyond that, the transformed text embedding matrix $\mathcal{A}(\mathbf{E})$ is inherently better suited for DDIM-based generative recommenders.
On the one hand, $\text{Var}(\mathcal{A}(\mathbf{E}))$ can better perverse variance in the DDIM framework. Taking $\mathcal{A}(\mathbf{e}) = \mathbf{e}\mathbf{O}\Lambda^{-\frac{1}{2}}$ for example:
\begin{align}
    \text{Var}(\mathcal{A}(\mathbf{E}))
    &= \Lambda^{-\frac{1}{2}}\mathbf{O}^T\text{Var}(\mathbf{E})\mathbf{O}\Lambda^{-\frac{1}{2}}  \\
    &= \Lambda^{-\frac{1}{2}}\mathbf{O}^T\mathbf{O} \Lambda\mathbf{O}^T\mathbf{O}\Lambda^{-\frac{1}{2}} = \mathbf{I}_d.
\end{align}
Referring to the forward Equation \eqref{eq:perturbe}, \ie $\mathbf{e}_L^t = \sqrt{\alpha_t} \mathbf{e}_L^0 + \sqrt{1 - \alpha_t} \bm{\epsilon}, \bm{\epsilon} \sim \mathcal{N}(\mathbf{0}, \mathbf{I})$, the noisy variance matrixe holds the following equation:
\begin{equation}
    \label{eq:var}
    \text{Var}(\mathbf{E}^t) := \alpha_t\text{Var}(\mathbf{E}) +(1-\alpha_t)\mathbf{I}_d.
\end{equation}
where $\text{Var}(\mathbf{E}^t)$ equals to $\mathbf{I}_d$ if $\text{Var}(\mathbf{E}) = \mathbf{I}_d$. Thus, $\text{Var}(\mathcal{A}(\mathbf{E})) \equiv\mathbf{I}_d$.
On the other hand, the transformed text embedding matrix $\mathcal{A}(\mathbf{E})$ will be more evenly dispersed in $\mathbb{R}^d$. The matrix $\mathbf{O}$ from Equation \eqref{eq:Osim} forms an orthonormal basis in $\mathbb{R}^d$ as long as the matrix $\mathbf{E}$ is full column rank \cite{LinearA1, LinearA2, PCA}. As a result, the row vectors of $\mathcal{A}(\mathbf{E})$ will be maximally separated from each other, collectively filling the space of $\mathbb{R}^d$ as fully as possible \cite{LinearA1, LinearA2, PCA}. This makes it easier for the generated oracle item embeddings to find the most relevant items.




\begin{figure*}[t!]
    \begin{subfigure}{0.48\linewidth}
        \centering
        \includegraphics[width=0.85\linewidth]{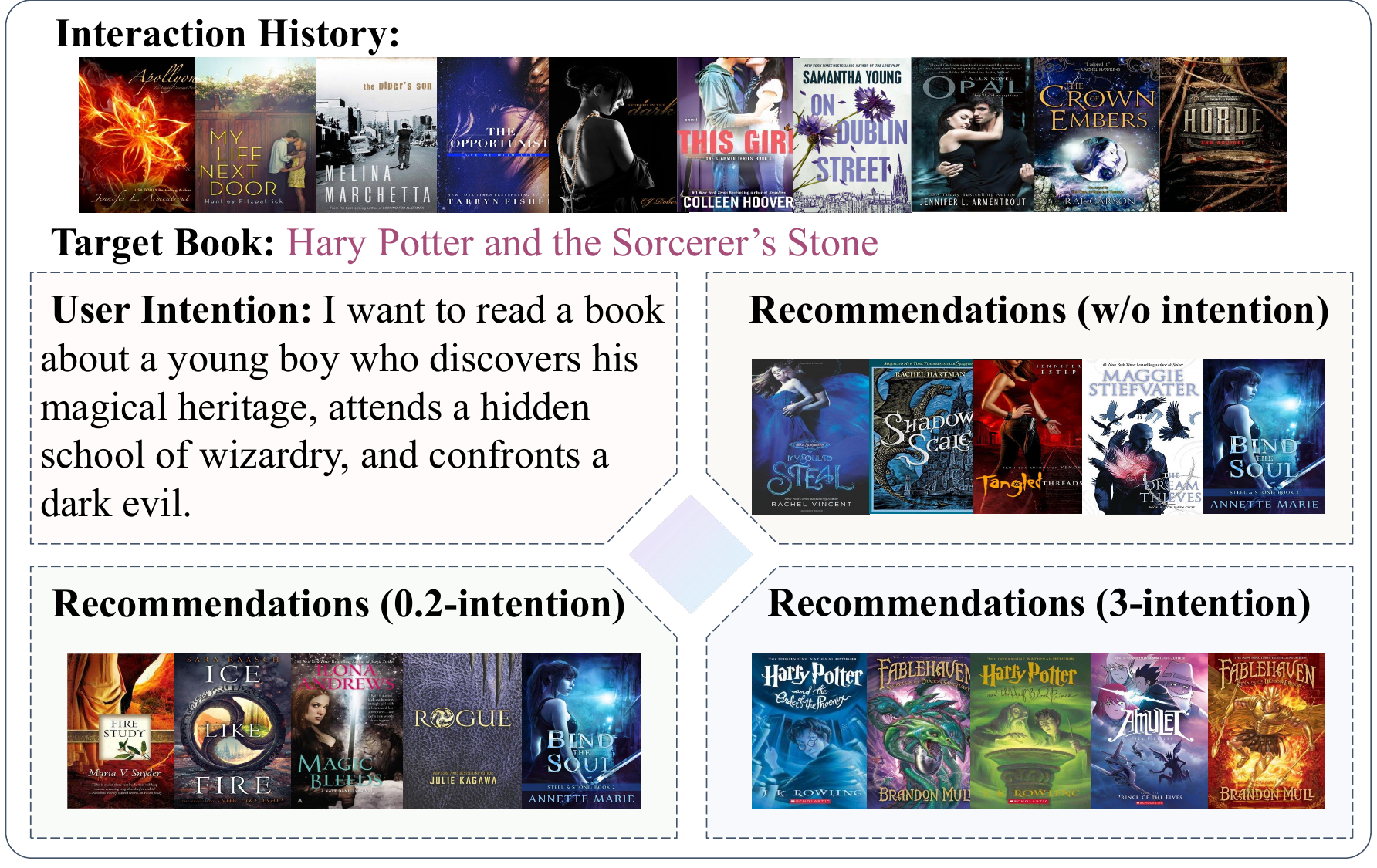}
        \label{case1}
    \end{subfigure}
    \begin{subfigure}{0.48\linewidth}
        \centering
        \includegraphics[width=0.85\linewidth]{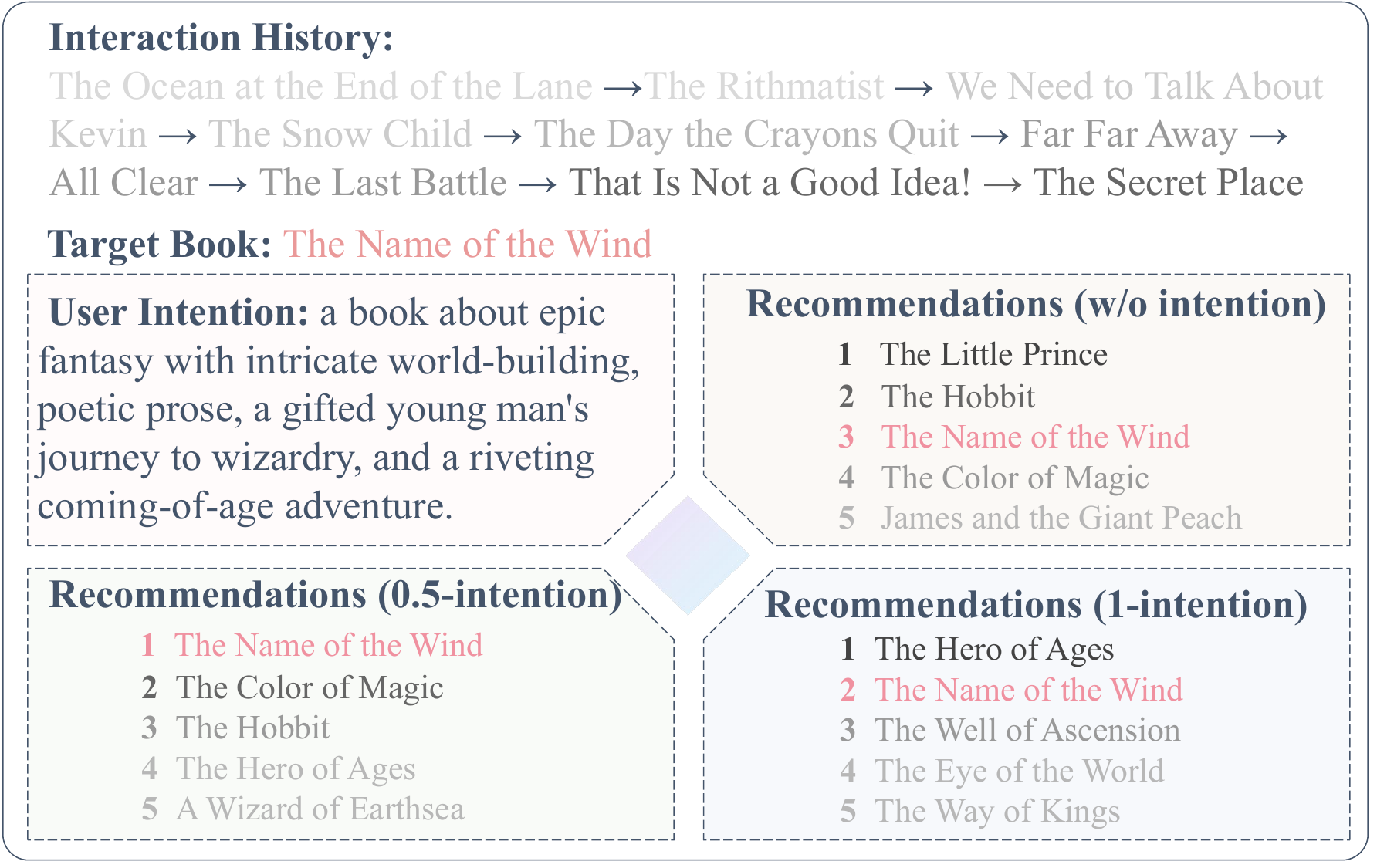}
        \label{case2}
    \end{subfigure}
    \caption{Two cases of intention guidance. $k$-intention indicates that the intention strength $\rho$ in Equation \eqref{mixG} is set to $k$.}
    \label{figCase}
\end{figure*}

\subsection{Intention Guidance via Text Embedding Models}\label{intentionG} \label{sec:method_intention}







As discussed in Section \ref{sec:intro}, another key limitation of current diffusion-based generative recommenders is the difficulty in incorporating more informative control signals, such as intention instructions, to guide the generation of oracle items.
This is largely due to the fact that the randomly initialized item embeddings are hardly interpretable by humans \cite{CDM, CDMT}.
In contrast, we have introduced TEM to acquire item embedding from textual descriptions of items, making them easier to understand in natural language.
This allows us to incorporate intention instructions, which reflect user intentions or prior knowledge, into the control signal.

Specifically, consider an intention instruction denoted as $\mathcal{Y}$ in natural language. 
Here is an example of intention instruction in the movie recommendation scenario: 
\begin{tcolorbox}[boxrule=0.5pt, left=0pt, right=0pt, top=2.5pt, bottom=2.5pt]
        \textit{A young boy discovers his magical heritage, attends a hidden school of wizardry, and confronts a dark evil.}
\end{tcolorbox}
By employing $\textbf{TEM-Enc}(\cdot)$, the encoder of TEM on the top of $\mathcal{Y}$, we can obtain the embedding of intention instruction $\mathcal{Y}$:
\begin{equation}
    \mathbf{e}_{\mathcal{Y}} = \textbf{TEM-Enc}(\mathcal{Y}),
\end{equation}
where $\mathbf{e}_{\mathcal{Y}}$ is the embedding of intention instruction, which lies in the same semantic space as the text embedding matrix of items $\mathbf{E}$ due to the employ of the same TEM encoder.

To generate oracle items align with these intention instructions, we can simply modify the generation process in iDreamRec by fusing $\mathbf{e}_{\mathcal{Y}}$ into the condition signal $\mathbf{e}_{<L}$ from interaction histories:
\begin{equation}
    \label{mixG}
    \mathbf{c} = \frac{1}{1+\rho}\left[\textbf{Cond-Enc}(\mathbf{e}_{<L}) + \rho\textbf{Cond-Enc}(\mathbf{e}_{\mathcal{Y}})\right],
\end{equation}
where $\textbf{Cond-Enc}(\cdot)$ is a module within the \textit{Denoiser} with arbitary architecture, and $\rho$ is a scalar representing the strength of intention instructions. In our experiments, we use a Diffusion Transformer (DiT) Block with in-context conditioning \cite{DiT}.

Similar to Equation \eqref{eq:DDIM_sample}, the intention-driven generation process can be formulated as:
\begin{equation}\label{eq:denoise_one_step}
     \hat{\mathbf{e}}_{L}^s = \sqrt{\alpha_{s}}\tilde{\mathbf{e}}_{\theta}(\hat{\mathbf{e}}_{L}^t, t,\mathbf{c})
            +\sqrt{1-\alpha_{s}}\frac{\hat{\mathbf{e}}_{L}^t-\sqrt{\alpha_{s}}\tilde{\mathbf{e}}_{\theta}(\hat{\mathbf{e}}_{L}^t, t, \mathbf{c})}{\sqrt{\alpha_{s}}},
\end{equation}
where $\hat{\mathbf{e}}_{L}^T\sim \mathcal{N}(\mathbf{0},\mathbf{I}_d)$. Thus, we can recursively compute it to generate the oracle item embeddings $\hat{\mathbf{e}}_{L}^0$.

Note that we do not explicitly involve intention instructions during the training process. 
Since the embedding of intention instruction $\mathbf{e}_{\mathcal{Y}}$ and the item embedding $\mathbf{e}_{v}$ are distributed in the same semantic space as the output space of TEM encoder, the intention instruction can be seamlessly integrated into the generation phase of iDreamRec as Algorithm \ref{alg:inference}, which can hardly be achieved in previous diffusion-based generative recommenders like DreamRec.

\subsection{Recommendation Strategies under Various Linear Transformations}

We are now capable of generating oracle item embeddings based on both interaction histories and intention instructions.
However, as observed in previous work \cite{DreamRec}, the generated oracle items may fall outside the candidate set.
Therefore, we need recommendation strategies that adapt to linear transformations which preserve variance within DDIM, while facilitating the recommendation of actual items more effectively, as explained in Section \ref{Sec:LT}.

\subsubsection{\textbf{Intention-understanding Recommendation under Dot-product Preserving}} \label{LT-PDP}

When set as $\mathcal{A}(\mathbf{e})=a\mathbf{e}, \mathcal{A}(\mathbf{e})=a\mathbf{e}\mathbf{O}$ ($\mathbf{O}$ is defined in Equation \eqref{eq:Osim}), the linear transformation $\mathcal{A}$ preserves the order of dot product, \ie $\langle \mathcal{A}(\mathbf{x}), \mathcal{A}(\mathbf{y})\rangle \ge \langle \mathcal{A}(\mathbf{x}), \mathcal{A}(\mathbf{z})\rangle $ if $\langle \mathbf{x}, \mathbf{y}\rangle \ge \langle \mathbf{x},\mathbf{z}\rangle $ for any $\mathbf{x},\mathbf{y},\mathbf{z}$. 
If transformation $\mathcal{A}(\cdot)$ preserves the order of the dot product, the diffusion model can learn the dot product structure during training, enabling it to generate oracle item embeddings guided by user textual intention.
Consequently, by recommending items that have the highest dot-product value with oracle item embedding, iDreamRec could retrieve real items in the candidate set under the guidance of both interaction histories and textual intention, as discussed in Section \ref{intentionG}. 

In addition to preserving dot-product, transformation $\mathcal{A}(\cdot)$ also maintains the mean of the embeddings, either as $\bm{\mu}=a\text{Mean}(\mathbf{E})$ or $\bm{\mu}=a\text{Mean}(\mathbf{E}\mathbf{O})$, implying the existence of distance between oracle item distribution $q(\mathbf{e}_L|\mathbf{e}_{<L})$ and pure noise $\mathcal{N}(\mathbf{0},\mathbf{I})$.
During the inference process, the generation of oracle items could be achieved by gradually recovering oracle item embedding from pure noise as Equation \eqref{eq:DDIM_sample}. 
Therefore, more inference steps are required to preserve the order of the dot product and bridge the distance between the oracle item distribution and pure noise.

\subsubsection{\textbf{Efficient Recommendation under Zero-mean Preserving}} \label{LT-PZM}

As analyzed above, we could set the linear transformation as
$\mathcal{A}(\mathbf{e})=(\mathbf{e}-\bm{u})\mathbf{O}\mathbf{\Lambda^{-\frac{1}{2}}},(\mathbf{e}-\bm{u})\mathbf{O}\bm{\Lambda}^{-\frac{1}{2}}\mathbf{O}$ and $(\mathbf{e}-\bm{u})\mathbf{O}\bm{\Lambda}^{-\frac{1}{2}}\mathbf{O}^T$ ($\bm{\mu} = \text{Mean}(\mathbf{E})$ and $\mathbf{O},\bm{\Lambda}$ are defined in Equation \eqref{eq:Osim}). Under these transformations, we have: $\text{Mean}(\mathcal{A}(\mathbf{e}_L^t)) = \sqrt{\alpha_t} \text{Mean}(\mathbf{e}_L^0) = \mathbf{0}$ and $\text{Var}(\mathcal{A}(\mathbf{e}_L^t)) = \alpha_t \text{Var}(\mathbf{e}_L^0) + (1-\alpha_t)\mathbf{I} = \mathbf{I}$. 
This implies that all noisy data distributions $\mathbf{e}_L^t$ have zero mean and an identity covariance matrix for all $t \in [0, T]$, indicating there is no average distance between them.
Under such circumstances, the inference process could be much more efficient in accuracy and time. Specifically, high recommendation accuracy can be achieved with a single-step generation, while multi-step generation further enhances it.




\section{Experiments}

In this section, we conducted extensive experiments to answer the following research questions: 
\begin{itemize}[leftmargin=*]
    \item \textbf{RQ1:} Could iDreamRec bring performance improvement by incorporating consistent item distribution modeling with Text Embedding Models?
    \item \textbf{RQ2:} Could iDreamRec generate oracle items under the guidance of intention instructions?
    \item \textbf{RQ3:} What are the main factors that impact the performance of iDreamRec?
\end{itemize}

\subsection{Experimental Setup}

\textbf{Datasets.} We use four real-world datasets:
\begin{itemize}[leftmargin=*]
    \item \textbf{Goodreads} is collected from the Goodreads platform \footnote{\url{https://www.goodreads.com/}}. We utilize ChatGPT to generate missing genres and descriptions.
    \item \textbf{MovieLens \cite{MovieLens}} is a commonly used movie recommendation dataset that contains user ratings, movie titles, and genres. For MovieLens, we utilize ChatGPT to generate descriptions.
    \item \textbf{Steam \cite{SASRec}} encompasses user reviews for video games on the Steam Store. For Steam, we utilize ChatGPT to generate descriptions.
    \item \textbf{Amazon-TV \cite{Amazon}} contains information and reviews for a wide range of movies on Amazon. Since the textual information is sufficiently rich, we do not prompt ChatGPT for any additional textual content.
\end{itemize}
In general, the four datasets represent different recommendation scenarios, while Amazon TV is more sparse, featuring over 40,000 items. The statistical information is shown in Table \ref{stat_dataset}. 
We retain the 20-core datasets and filter out users and items with fewer than twenty interactions except MovieLens, which keeps the original dataset. 
Subsequently, we arrange all sequences chronologically and preserve the last 10 interactions as interaction history. 
For sequences with less than 10 interactions, we would pad them to 10 with a padding token. 
Then, we partition the data into training, validation, and testing sets at an 8:1:1 ratio. 
Regarding item text, we utilized GPT to enhance the textual information of the items, except for Amazon TV, which has sufficiently rich text information. 
Then, we concatenate the text information of items, including titles, genres, and descriptions, and input them into OpenAI's TEM \textit{text-embeddings-3-small} to obtain textual item embeddings. For statistics of datasets, please refer to Table \ref{stat_dataset}.

\begin{table*}[t]
    \centering
    \caption{Overall Performance comparison. Bold and underlined respectively indicate the best and the second-best performance. The experiments are conducted 10 times and the average results are reported. ``*'' denotes that the improvements are significant using $t$-test for 10 samples with $p$-value $< 0.01$ .}
    \label{main experiment}
    \renewcommand{\arraystretch}{1.2} 
    \vspace{-5pt}
    \resizebox{\linewidth}{!}{
    \begin{tabular}{c|cccc|cccc|cccc|cccc}
    \toprule
    \multirow{2}{*}{Models} & \multicolumn{4}{c|}{Goodreads} & \multicolumn{4}{c|}{MovieLens} & \multicolumn{4}{c|}{Steam} & \multicolumn{4}{c}{Amazon-TV} \\
    & HR@5 & NDCG@5 & HR@10 & NDCG@10 & HR@5 & NDCG@5 & HR@10 & NDCG@10 & HR@5 & NDCG@5 & HR@10 & NDCG@10 & HR@5 & NDCG@5 & HR@10 & NDCG@10\\
    \midrule
    Caser  
    &0.0909&0.0619&0.1298& 0.0744
    &\underline{0.0934}&\underline{0.0588}&0.1526&\underline{0.0778} 
    &0.0214&0.0138&0.0359&0.0185 
    &0.0100&0.0060&0.0175&0.0084
    \\
    SASRec 
    &0.1201  &0.0825&0.1386& 0.0885 
    &0.0777&0.0392&0.1457&0.0611
    &\underline{0.0362}&\underline{0.0217}&0.0538&\underline{0.0273} 
    &\underline{0.0193}&\underline{0.0112}&\underline{0.0318} &\underline{0.0153}
    \\
    \midrule
    MoRec 
    & 0.1146 & 0.0642 & \underline{0.1442} & 0.0738 
    & 0.0770 & 0.0382 & 0.1492 & 0.0614 
    & 0.0262 & 0.0134 & 0.0452 & 0.0195 
    & 0.0102 & 0.0073 & 0.0132 & 0.0082\\
    UniSRec 
    &\underline{0.1205}&\underline{0.1032}&0.1318& \underline{0.1069} 
    &0.0812&0.0403  &\underline{0.1582}&0.0650  
    &0.0338 &0.0173& \underline{0.0640}&0.0270 
    &0.0101&0.0080&0.0120&0.0086
    \\
    \midrule
    DiffRec  
    &0.0346&0.0170&0.0861&0.0335 
    &0.0252&0.0113&0.0901&0.0322
    &0.0240&0.0134&0.0559&0.0235 
    &0.0118&0.0063&0.0217&0.0095
    \\
    DreamRec 
    &0.1101&0.0968&0.1179&0.0993 
    &0.0889&0.0577& 0.1372   &0.0733 
    & 0.0074& 0.0059&0.0098&0.0066 
    &0.0036&0.0033&0.0040&0.0035
    \\
    \midrule
    iDreamRec 
    &$\textbf{0.1387}^{*}$&$\textbf{0.1124}^{*}$&$\textbf{0.1611}^{*}$&$\textbf{0.1196}^{*}$ 
    & $\textbf{0.1217}^{*}$& $\textbf{0.0782}^{*}$& $\textbf{0.1929}^{*}$& $\textbf{0.1011}^{*}$ 
    & $\textbf{0.0500}^{*}$& $\textbf{0.0358}^{*}$& $\textbf{0.0715}^{*}$& $\textbf{0.0427}^{*}$
    &$\textbf{0.0252}^{*}$&$\textbf{0.0181}^{*}$&$\textbf{0.0357}^{*}$&$\textbf{0.0215}^{*}$
    \\
    Improv. 
    & 15.14\%&8.84\%&11.72\%&11.92\% 
    &30.37\%&32.08\%&21.91\%&30.03\% 
    & 38.12\%& 65.16\%& 11.67\%& 58.35\%
    &30.32\%&60.58\%&12.26\%&40.52\%
    \\
    \bottomrule
    \end{tabular}}
\end{table*}

\begin{figure*}[t!]
    \begin{subfigure}{0.47\linewidth}
        \centering
        \includegraphics[width=0.9\textwidth]{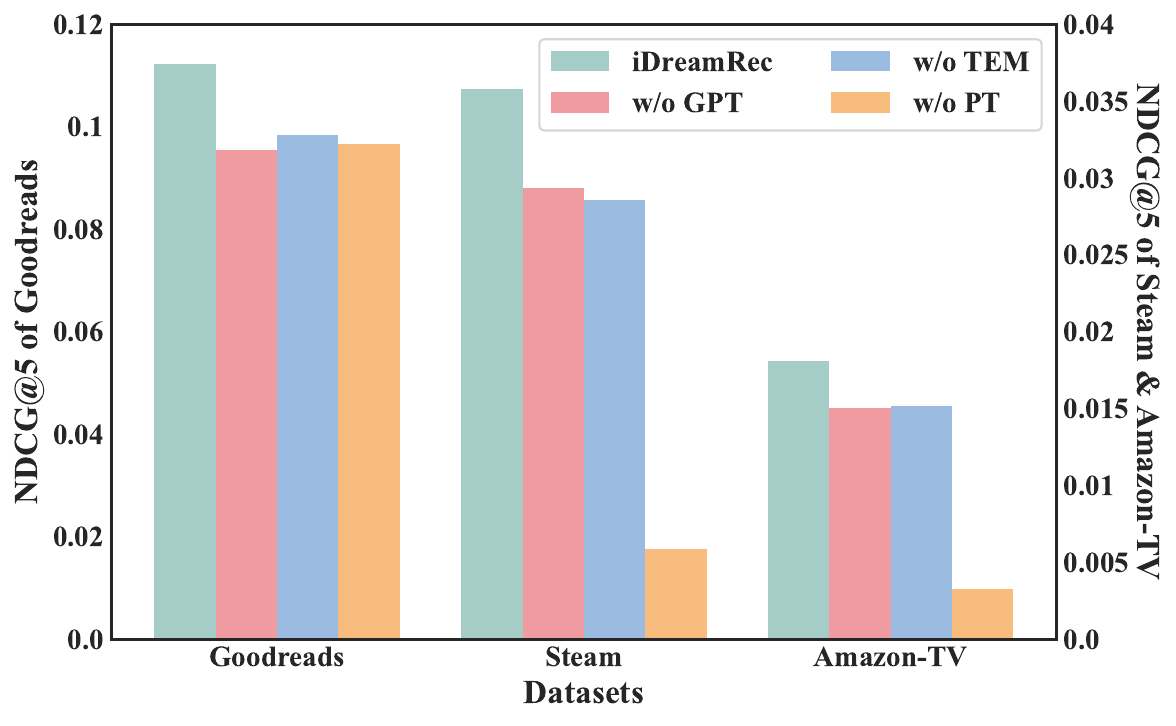}
        \caption{Abalation on item embeddings}
        \label{fig:aba_emb}
    \end{subfigure}
    \begin{subfigure}{0.47\linewidth}
        \centering
        \includegraphics[width=0.9\textwidth]{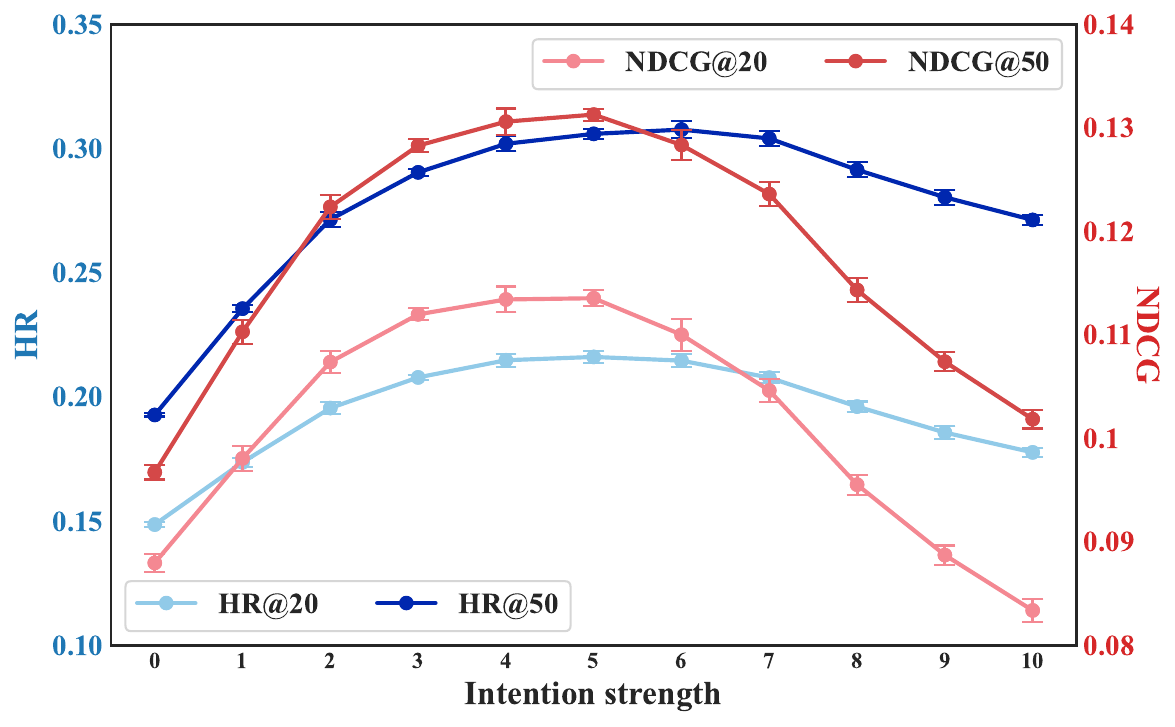}
        \caption{Results of intention Guidance on Goodreads}
        \label{fig:exp_intention}
    \end{subfigure}
    \caption{The left figure shows the performance of iDreamRec under different datasets with varying types of item embeddings: 1) \textbf{iDreamRec:} text embeddings are derived as the method. 2) \textbf{iDreamRec w/o GPT:} text embeddings are merely derived from the name of items. 3) \textbf{iDreamRec w/o TEM:} iDreamRec with pretrained ID embeddings from a SASRec. 4) \textbf{iDreamRec w/o PT:} iDreamRec with learnable ID embeddings. The right figure is the results of intention guidance on Goodreads.}
    \label{figEmb}
\end{figure*}

\vspace{5pt}
\noindent\textbf{Compared Methods.}  We compare the proposed approach with the following baseline methods:
\begin{itemize}[leftmargin=*]
    \item \textbf{SASRec} \cite{SASRec} takes advantage of the self-attention mechanism in Transformer to capture users' historical preference from interaction histories using learnable ID embedding.
    \item \textbf{Caser} \cite{Caser} uses horizontal and vertical convolutional filters to capture sequential patterns at the point-level, union-level to allow skip behaviors using learnable ID embedding.
    \item \textbf{MoRec} \cite{MoRec} employs item characteristics derived from text descriptions encoded with a text encoder, or other modal information with its corresponding modal encoder, and performs dimensional transformation to align with SASRec.
    \item \textbf{UniSRec} \cite{UniSRec} proposes an architecture based on parametric whitening and a mixture-of-experts enhanced adaptor to transform original text embeddings to adapt SASRec.
    \item \textbf{DiffRec} \cite{DRM}, on behalf of diffusion-based discriminative recommendation models, utilizes diffusion models to model the distribution of probability vectors for each item per user.
    \item \textbf{DreamRec} \cite{DreamRec}, a diffusion-based generative recommender, reshape sequential recommendation as oracle item generation.
\end{itemize}


Overall, we compared traditional discriminative recommenders using trainable ID embeddings represented by SASRec and Caser, recommenders based on text embeddings represented by MoRec and UniSRec, and diffusion-based generative recommenders represented by DiffRec and DreamRec. 

To ensure a fairer comparison, the methods based on text embeddings all utilize the same text embeddings from OpenAI's TEM \textit{text-embeddings-3-small} as ours iDreamRec. 
After acquiring text embeddings by OpenAI's TEM \textit{text-embeddings-3-small}, iDreamRec utilize linear transformation $\mathbf{A}(\mathbf{e})=(\mathbf{e}-\bm{u})\mathbf{O}\bm{\Lambda}^{-\frac{1}{2}}\mathbf{O}^T$ or $\mathbf{A}(\mathbf{e})=(\mathbf{e}-\bm{u})\mathbf{O}\Lambda^{-\frac{1}{2}}$ to preserve zero-mean and variance of identity matrix as section \ref{LT-PZM}. 
Thereby, we could achieve one-step generative recommendation with high sampling quality.
In our experiments, UniSRec and MoRec also get use of the same text embeddings instead of \textbf{[CLS]} embeddings from BERT.


\vspace{5pt}
\noindent\textbf{Training Protocol.} We implement all models with Python 3.7 and PyTorch 1.12.1 in Nvidia GeForce
RTX 3090.  We leverage AdamW as the optimizer.
The embedding dimension of items is fixed as 1536 across all models. The learning rate is tuned in the
range of $\left\{0.001, 0.0001, 0.00001, 0.000001\right\}$. 
For SASRec, Caser, MoRec, and UniSRec, we use \textit{binary cross-entropy} loss and perform negative sampling randomly at the ratio of $\left\{16,32,64\right\}$, while there is no negative sampling in DiffRec, DreamRec, and our iDreamRec. For both DreamRec and our iDreamRec, we fix the unconditional
training probability $p$ as 0.1 suggested by \cite{CFG} and set the total diffusion step $T$ fixed as 2000, and the personalized guidance strength $w$ in the range of $\left\{0, 2, 4, 6, 8, 10\right\}$ as DreamRec. As for inference, DreamRec conducts $T$-step generation while our iDreamRec achieves one-step generation.

\vspace{5pt}
\noindent\textbf{Evaluation Protocol. } We adopt two widely-used metrics to evaluate the performance
of recommendation \cite{SASRec, DreamRec}: hit ratio (HR$@K$) and normalized
discounted cumulative gain (NDCG$@K$), where $K$ is set to 5 and 10. And we calculate them based on the rankings of all items other than some sampled items. The full ranking strategy to some extent provides a fairer assessment \cite{full-rank}.

\begin{figure*}[t] 
    \centering 
    \includegraphics[width=0.9\textwidth]{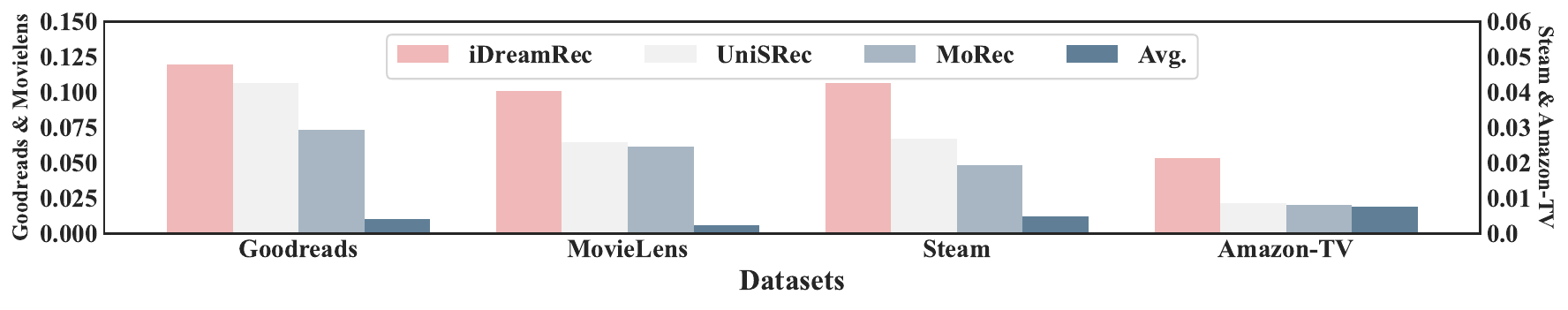}
    \vspace{-5pt}
    \caption{Analyzing the performance (NDCG@10) of different recommenders with text embeddings. Avg. refers to the case where we compute the cosine similarity between the target item embeddings and the mean of the text embeddings in the interaction history, as indicated by logic. }
    \label{figText}
    \vspace{-5pt}
\end{figure*}

\begin{figure*}[t] 
    \centering 
    \includegraphics[width=0.9\textwidth]{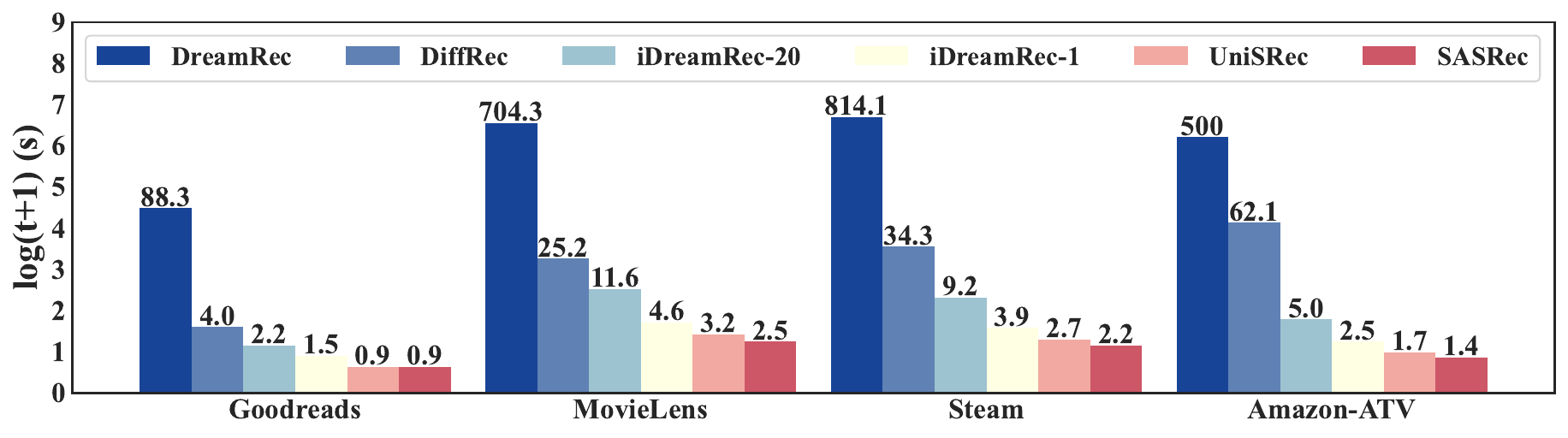}
    \vspace{-5pt}
    \caption{Analyzing the inference efficiency of iDreamRec compared to baseline methods. The inference time $t$ seconds represents the time taken by the model to complete inference on the test set, and it is indicated above the bars in the bar chart. The height of each bar corresponds to $\log(t+1)$. iDreamRec-20 indicates a 20-step inference as described in Algorithm \ref{alg:inference}.}
    \label{figTime}
    \vspace{-5pt}
\end{figure*}

\subsection{Main Results (RQ1)} 
In this section, we compare iDreamRec with the baselines in terms of recommendation performance and analyze the reasons behind the results.
The main experiment results are shown in Table \ref{main experiment}.

In general, iDreamRec substantially outperforms compared approaches, highlighting the effectiveness of consistent item embedding modeling for diffusion-based generative recommenders, which is achieved by prompting ChatGPT and integrating pretrained Text Embedding Models in iDreamRec.
Moreover, we can observe that DreamRec achieves pleasing performance on MovieLens and Goodreads data, while its performance on Steam and Amazon-TV is unsatisfactory.
This performance gap can be attributed to that the inconsistency of randomly initialized ID embeddings would be amplified with the growth in the number of items, with much more items in Steam and Amazon-TV data than MovieLens and Goodreads.
In contrast, iDreamRec models a more consistent embedding distribution of items, thus maintaining the best performance with a larger number of items. 



\subsection{Results with Intention Instructions (RQ2)}

In this section, we progress to further demonstrate the intention instruction following the capability of iDreamRec, where a practical challenge lies in the lack of intention information in public recommendation datasets.
To resolve this challenge, we compromise to prompt ChatGPT to generate the user intention $\mathcal{Y}$ by providing a textual description of the item of interest to the user.
Afterward, we examine whether iDreamRec could follow intention instruction $\mathcal{Y}$ and provide higher recommendation accuracy towards the target item as discussed in Section \ref{sec:method_intention}.
The results are shown in Figure \ref{figIntention}.

From Figure \ref{figIntention} we can observe that by varying the guidance strength of intention instructions, iDreamRec shows sensitiveness towards the prediction accuracy on target items, and could outperform historical interaction guidance under certain guidance strengths.
Note that the intention instruction is not involved in the training of iDreamRec, and only integrated in the generation phase as described in Section \ref{sec:method_intention}, which implies the flexibility and stability of integrating more text-based guidance in iDreamRec.

\subsection{Ablation Studys (RQ3)}

To answer RQ3, we conducted ablation experiments to deeply analyze the contribution of key considerations in iDreamRec.

Firstly, we conduct an ablation experiment to analyze the impact of different types of embeddings on iDreamRec. 
The results are shown in Figure \ref{figEmb}.
To sum up, \textbf{the consistency of item embeddings is a crucial adjective} of the performance of iDreamRec, even diffusion-based generative recommenders. 
And \textbf{rich item textual information will also be beneficial} to the performance of iDreamRec with text embeddings. 
Moreover, even when using pre-trained ID embeddings, iDreamRec still demonstrates good performance, indicating that other types of consistent item embeddings are also applicable to iDreamRec.

Secondly, we conduct a study on the impact of recommender frameworks on text embeddings, with the results presented in Figure ref{figText}.
The results demonstrate that \textbf{diffusion-based recommenders can leverage consistent text embeddings from TEM more effectively than traditional recommenders}. 

\subsection{Time Efficiency}

The high inference time, a well-known drawback of diffusion models, renders diffusion-based recommenders impractical. Therefore, we will analyze the inference time efficiency of iDreamRec in this section.
The results are shown in Figure \ref{figTime}.
As shown in the figure, iDreamRec significantly improves inference efficiency compared to other diffusion-based recommendation systems, and it is even comparable to traditional recommenders.


\section{Conclusion}

We propose iDreamRec, which models consistent, context-aware embeddings encoded by advanced Text Embedding Models (TEM) and incorporates intention instructions for guiding the generation of the oracle item.
Specifically, iDreamRec substantially encapsulates prior knowledge encoded in LLMs and TEMs into fixed item embeddings, thus nudging diffusion models to capture a more consistent oracle item distribution than ID embeddings.
More importantly, the item embedding space of iDreamRec lies in accordance with the output space of TEM, allowing for the use of more informative intention instructions for oracle item generation.
Experiments demonstrate that iDreamRec consistently enhances diffusion-based generative recommenders and enables the integration of intention instructions for more precise and effective recommendation results.




\section{Limitation}
There are also a few limitations of iDreamRec: 
\begin{itemize}[leftmargin=*]
    \item \textbf{underexplored intention instructions}: The intention instructions in iDreamRec are either heuristically constructed or generated by GPT, which differ from actual user inputs. It arises from the lack of real-world intention instruction data. Additionally, the effectiveness of intention guidance based on Text Embedding Models (TEM) is worth exploring. Our experiments revealed that, apart from the semantic of intentions, the syntactic structure might also has an impact as shown in Figure \ref{figCase}.
    \item \textbf{high latent dimension}: The latent dimension of iDreamRec is determined by the output dimension of TEM, which is often significantly larger than the hidden dimension of traditional recommendation systems, typically set at 64. 
\end{itemize}

\newpage

\bibliographystyle{ACM-Reference-Format}
\bibliography{ref}

\appendix

\newpage

\appendix

\section{Denosing Diffusion Implicit Models} \label{DDIM}

\subsection{Simplify ELBO} \label{A.1}

Given samples from a consistent data distribution $q(\mathbf{x}_0)$, we aim to learn a model distribution
$p_{\theta}(\mathbf{x}_0)$ that approximates $q(\mathbf{x}_0)$ and is easy to sample from. DDIM are latent variable models of the form
\begin{equation}
\label{eq1}
p_{\theta}(\mathbf{x}_0) = \int p_{\theta}(\mathbf{x}_{0:T}) \mathrm{d}\mathbf{x}_{1:T}, \ \text{where}\ p_{\theta}(\mathbf{x}_{0:T})=p_{\theta}(\mathbf{x}_{T})\sum\limits_{t=1}^{T}p_{\theta}(\mathbf{x}_{t-1}|\mathbf{x}_t)
\end{equation}
where $\mathbf{x}_1, \dots,\mathbf{x}_T$ are noisy variables of the same dimension $d$ as $\mathbf{x}_0$ (the space denoted as $\mathcal{X}$ ). The parameters $\theta$ are learned to fit the data distribution $q(\mathbf{x}_0)$ by maximizing the evidence lower bound of the log-likelihood.
\begin{align}
    \label{eq2}
    \log p_{\theta}(\mathbf{x}_0)
    &= \log \mathbb{E}_{q(\mathbf{x}_{1:T}|\mathbf{x}_{0})}\left[\frac{p_{\theta}(\mathbf{x}_{0:T})}{q(\mathbf{x}_{1:T}|\mathbf{x}_0)}\right] \\
    &\ge \mathbb{E}_{q(\mathbf{x}_{1:T}|\mathbf{x}_{0})}\left[\log \frac{p_{\theta}(\mathbf{x}_{0:T})}{q(\mathbf{x}_{1:T}|\mathbf{x}_0)}\right] = \text{ELBO} 
\end{align}
We assume that the noisy variable $\mathbf{x}_t$  follows the forward equation $q(\mathbf{x}_t|\mathbf{x}_0) := \mathcal{N}(\mathbf{x}_t;\sqrt{\alpha_t}\mathbf{x}_0,(1-\alpha_t)\mathbf{I}_d)$ to persevere the variance as $\text{Var}(\mathbf{x}_t)=\alpha_t\text{Var}(\mathbf{x}_0) + \left(1-\alpha_t\right)\mathbf{I}_d.$ Moreover, we could simplify ELBO as follows. 
\begin{align}
\label{eq3}
\text{ELBO} 
&=\underbrace{\mathbb{E}_{q(\mathbf{x}_1|\mathbf{x}_0)}\left[\log p_{\theta}(\mathbf{x}_{0}|\mathbf{x}_1)\right]}_{\mathcal{L}_{0}} \\
&+\sum\limits_{t=2}^{T}\underbrace{\mathbb{E}_{q(\mathbf{x}_t|\mathbf{x}_0)}\left[D_{\text{KL}}(q(\mathbf{x}_{t-1}|\mathbf{x}_t,\mathbf{x}_0)||p_{\theta}(\mathbf{x}_{t-1}|\mathbf{x}_t)\right]}_{\mathcal{L}_{t-1}} +\text{C}
\end{align}
\begin{equation}
\label{eq6}
\mathcal{L}_{t}=\mathbb{E}_{q(\mathbf{x}_{t+1}|\mathbf{x}_0)}\left[\frac{\alpha_{t}}{2(1-\alpha_{t})^2}\Vert \hat{\bm{x}}_{\theta}(\mathbf{x}_{t+1},t+1)-\mathbf{x}_0 \Vert_2^2\right]
\end{equation}

where $\hat{\bm{x}}_{\theta}$ is the output of the network with arbitrary architecture, named \textit{Denoiser}, to predict $\mathbf{x}_0$.

\subsection{Accelerated Sampling Processes.} \label{A.2}
Given that $\mathbf{x}_t = \sqrt{\alpha_t}\mathbf{x}_0 + \sqrt{1-\alpha_t}\bm{\epsilon}, \bm{\epsilon}\sim\mathcal{N}(\mathbf{0},\mathbf{I}_d)$ holds for any integer $t\in \left[1,\dots,T\right]$, we could also assume that the transition probability $q(\mathbf{x}_t|\mathbf{x}_s,\mathbf{x}_0) \ \text{with}\ s<t$ follows Gaussian distribution:
\begin{equation}
\label{eq8}
q(\mathbf{x}_t|\mathbf{x}_s,\mathbf{x}_0) := \mathcal{N}(\mathbf{x}_t; \sqrt{\alpha_t} \mathbf{x}_0 + \sqrt{1-\alpha_t-\sigma_t^2}\frac{\mathbf{x}_s-\sqrt{\alpha_s}\mathbf{x}_0}{\sqrt{1-\alpha_s}},\sigma_t^2\mathbf{I}_d)
\end{equation}


Referring to the Bayesian formula, we can derive the reverse transition probability distribution $q(\mathbf{x}_s|\mathbf{x}_t,\mathbf{x}_0)\ \text{with}\ s<t$ as follows:
\begin{equation}
\label{eq9}
q(\mathbf{x}_s|\mathbf{x}_t,\mathbf{x}_0) = \mathcal{N}(\mathbf{x}_s; \sqrt{\alpha_s} \mathbf{x}_0 + \sqrt{1-\alpha_s-\sigma_s^2}\frac{\mathbf{x}_t-\sqrt{\alpha_t}\mathbf{x}_0}{\sqrt{1-\alpha_t}},\sigma_s^2\mathbf{I}_d)
\end{equation}

For any $s<t$, there is Equation \eqref{eq10} with $\sigma_s =0$ which means that we can train a model with an arbitrary number of forward steps $T$ but only
sample by a subsquence with a length less than $T$ in the inference process (\eg $\text{let}\ s=t-100$ leading to $\frac{T}{100}$ sampling steps in stead of $T$).

\begin{equation}
\label{eq10}
 \mathbf{x}_s = \sqrt{\alpha_s} \hat{\bm{x}}_{\theta}(\mathbf{x}_t, t) + \sqrt{1-\alpha_s}\frac{\mathbf{x}_t-\sqrt{\alpha_t}\hat{\bm{x}}_{\theta}(\mathbf{x}_t, t)}{\sqrt{1-\alpha_t}} 
\end{equation}

\subsection{Modeling the gradients of conditional likelihood.} \label{A.3}
Mathematically, for a Gaussian variable $\mathbf{z}\sim\mathcal{N}(\mathbf{z};\bm{\mu}_z, \mathbf{\Sigma}_z)$, Tweedie's Formulation states that:

\begin{equation}
\label{eq11}
\mathbb{E}\left[\bm{\mu}_z|\mathbf{z}\right] = \mathbf{z} + \mathbf{\Sigma}_z\nabla_{\mathbf{z}}\log q(\mathbf{z}). 
\end{equation}

Consider that $\mathbf{x}_t \sim \mathcal{N}(\mathbf{x}_t;\sqrt{\alpha_t}\mathbf{x}_0, \left(1-\alpha_t\right)\mathbf{I}_d)$, there are $\sqrt{\alpha_t}\mathbf{x}_0 = \mathbf{x}_t + (1-\alpha_t)\nabla_{\mathbf{x}_t}\log q(\mathbf{x}_t)$. 

As for approximate data distribution $p_{\theta}(\mathbf{x}_t)$ towards $q(\mathbf{x}_t)$, there holds:

\begin{equation}
\label{eq12}
\hat{\bm{x}}_{\theta}(\mathbf{x}_{t},t) = \frac{1}{\sqrt{\alpha_t}}\mathbf{x}_t + \frac{(1-\alpha_t)}{\sqrt{\alpha_t}}\nabla_{\mathbf{x}_t}\log p_{\theta}(\mathbf{x}_t),
\end{equation}
where direction $\nabla_{\mathbf{x}_t}\log p_{\theta}(\mathbf{x}_t)$ measures how to move in data space $\mathbb{R}^{d}$ to maximize Equation \eqref{eq2}. 

To achieve controllable generation, we need to model conditional distribution under condition $\mathbf{c}$:
\begin{align}
    \label{eq13}
    \nabla_{\mathbf{x}_t}\log p_{\theta}(\mathbf{x}_t|\mathbf{c}) 
    &= \nabla_{\mathbf{x}_t}\log \left(\frac{p_{\theta}(\mathbf{x}_t)p_{\theta}(\mathbf{c}|\mathbf{x}_t)}{p_{\theta}(\mathbf{c})}\right) \\
    \label{eq14}
    &= \nabla_{\mathbf{x}_t}\log p_{\theta}(\mathbf{x}_t) + \nabla_{\mathbf{x}_t}\log p_{\theta}(\mathbf{c}|\mathbf{x}_t)
\end{align}

According to Equation \eqref{eq14}, we estimate the gradients of conditional likelihood $\nabla_{\mathbf{x}_t}\log p_{\theta}(\mathbf{c}|\mathbf{x}_t) = w\left(\nabla_{\mathbf{x}_t}\log p_{\theta}(\mathbf{x}_t|\mathbf{c}) - \nabla_{\mathbf{x}_t}\log p_{\theta}(\mathbf{x}_t)\right)$ scaling with condition strength parameter $w$.

In conclusion, we adjust the output of \textit{Denoiser} as Equation \eqref{eq14} to achieve conditional generation.

\begin{align}
    \tilde{\bm{x}}_{\theta}(\mathbf{x}_{t},t,\mathbf{c}) 
    \label{eq15}
    &= \hat{\bm{x}}_{\theta}(\mathbf{x}_{t},t, \mathbf{c}) + w\left(\nabla_{\mathbf{x}_t}\log p_{\theta}(\mathbf{x}_t|\mathbf{c}) - \nabla_{\mathbf{x}_t}\log p_{\theta}(\mathbf{x}_t)\right)
\end{align}



\section{Algorithms and Case Studys} \label{alg}

\begin{table}[h!]
\centering
\caption{Symbol Definitions and Descriptions}
\label{tab:symbols}
\begin{tabular}{|c|l|}
\hline
\textbf{Symbol} & \textbf{Description} \\ \hline
$t$ & Diffusion time step \\ \hline
$\mathbf{v}_{<L}$ & Interaction history of items \\ \hline
$v_L$ & Target item \\ \hline
$\mathbf{e}_{<L}$ & Interaction history of item embeddings \\ \hline
$\mathbf{e}_L, \mathbf{e}_L^0$ & Embedding of target item \\ \hline
$t$ & Diffusion time step \\ \hline
$\alpha_t$ & Noise level at step $t$ \\ \hline
$\mathbf{e}_L^t$ & Noisy embedding at noise level $\alpha_t$ \\ \hline
$\mathbf{e}_L^{0:T}$ & joint variable of $\mathbf{e}_L^t,\forall t\in \{0,\cdots,T\}$ \\ \hline
$\mathcal{U}(a, b)$ & Uniform distribution in the range $[a, b]$ \\ \hline
$\mathcal{N}(\mu, \sigma^2)$ & Normal distribution with mean $\mu$ and variance $\sigma^2$ \\ \hline
$\mathbf{I}_d$ & Unit matrix \\ \hline
$\mathcal{Y}$ & User intention generated by the model \\ \hline
$\theta$ & Model parameters \\ \hline
$\mathcal{L}$ & Loss function \\ \hline
$\mathcal{D}$ & Dataset \\ \hline
\end{tabular}
\end{table}

\begin{algorithm}[H]  
  \caption{ Training Phase of iDreamRec}  
  \label{alg:train}  
  \begin{algorithmic}[t]
    \Require  
        uncondition probability $p$,
        noise schedule $\{\alpha_t\}_{t=1}^{T}$,
        dataset $\mathcal{D} = \{ \left(\mathbf{e}_{<L}, \mathbf{e}_L\right)_k \}_{k=1}^{|\mathcal{D}|}$.  
    \Ensure  
        optimal \textit{Denoiser} $\hat{\mathbf{e}}_{\theta}(\mathbf{e}_L^t, t, \mathbf{c})$ with uncondition vector $\Phi$.
    \Repeat  
        \State $\mathbf{e}_{<L}, \mathbf{e}_L \sim \mathcal{D}$  
        \State $t\sim \text{Uniform}(\{1,\dots,T\})$
        \Comment{sampling timestep $t$ uniformly.}
        \State $\mathbf{c}=\textbf{\text{Cond-Enc}}(\mathbf{e}_{<L})$ with probability $p$ or $C=\Phi $ \\ \Comment{for classifier-free guidance}
        \State $\bm{\epsilon} \sim \mathcal{N}(\mathbf{0},\mathbf{I}_d)$ \State$\mathbf{e}_L^t=\sqrt{\alpha_t}\mathbf{e}_L^0+\sqrt{1-\alpha_t}\bm{\epsilon}$
        \Comment{pertube $\mathbf{e}_L^0$}
        \State Take gradient descent step on
        $\nabla_{\theta}\Vert \mathbf{e}_L^0 - \hat{\mathbf{e}}_{\theta}(\mathbf{e}_L^t, t, \mathbf{c})\Vert_2^2$
    \Until{converged}
  \end{algorithmic} 
\end{algorithm}

\begin{algorithm}[H]  
  \caption{ Inference Phase of iDreamRec}  
  \label{alg:inference}  
  \begin{algorithmic}[t]
    \Require  
        user's intention embedding $\mathbf{e}_Y$, test dataset $\mathcal{D}_t = \{ \left(\mathbf{\mathbf{e}}_{<L}, \mathbf{e}_L\right)_k \}_{k=1}^{|\mathcal{D}_t|}$, a sequence $\tau=\{\tau_i\}_{i=1}^{|\tau|}$ with $\tau_S=T$ and $\tau_1=1$ for multi-step denoising in Equation \eqref{eq:DDIM_sample}.
    \Ensure
        oracle item embedding $\hat{\mathbf{e}}_L^0$
        \State $\mathbf{e}_{<L}, \mathbf{e}_L \sim \mathcal{D_t}$
        \State $\mathbf{c}_s=\textbf{\text{Cond-Enc}}(\mathbf{e}_{<L})$ as well as $\mathbf{c}_y=\textbf{\text{Cond-Enc}}(\mathbf{e}_y)$ 
        \State $\mathbf{c}=\frac{1}{1+\rho}(\mathbf{c}_s+\rho\mathbf{c}_y)$  \Comment{$\rho$ is intention strength}
        \State $\hat{\mathbf{e}}_T=\hat{\mathbf{e}}_{\tau_{|\tau|}}\sim \mathcal{N}(\mathbf{0},\mathbf{I}_d)$ 
        \For {$s=|\tau|$ to 1}
            \State $\tilde{\mathbf{e}}_{0} =(1+w)\hat{\mathbf{e}}_{\theta}(\hat{\mathbf{e}}_L^{\tau_s}, \tau_s, \mathbf{c}) - w\hat{\mathbf{e}}_{\theta}(\hat{\mathbf{e}}_L^{\tau_s}, \tau_s, \Phi)$ \\ \Comment{$w$ is conditional guidance strength} 
            \State $\hat{\mathbf{e}}_{\tau_{s-1}} = \sqrt{\alpha_{\tau_{s-1}}}\tilde{\mathbf{e}}_{0}
            +\sqrt{1-\alpha_{\tau_{s}}}\frac{\hat{\mathbf{e}}_{\tau_s}-\sqrt{\alpha_{\tau_{s-1}}}\tilde{\mathbf{e}}_{0}}{\sqrt{\alpha_{\tau_{s-1}}}}$
        \EndFor \\
        \Return $\hat{\bm{x}}_{0}$ \Comment{the oracle item embedding.}
  \end{algorithmic} 
\end{algorithm} 

\begin{table}[t]
	\caption{Statistics of datasets.}
	\label{stat_dataset}
	\centering
        \renewcommand\arraystretch{1.1}
	\begin{tabular}{llll}
		\toprule 
		Dataset  & \# users
		& \# items & sparsity 
		\\
        \midrule
         MovieLens
		& 6040   & 3883
		& 0.0427   \\
        Goodreads
		& 6032   & 4550
        & 0.0080 \\
         Steam
		& 39795   & 9265
		& 0.0042   \\
         Amazon-TV
		& 20515   & 44014
		& 0.0007   \\
        \midrule
        \end{tabular}
        \vspace{-5pt}
\end{table}

\end{document}